\documentclass[prb,
11pt,
superscriptaddress,showpacs,amsmath,amssymb]{revtex4}

\begin{document}

\author{G.E.~Volovik}
\affiliation{Landau Institute for Theoretical Physics, acad. Semyonov av., 1a, 142432,
Chernogolovka, Russia}

\title{Proton decay in de Sitter environment}

\date{\today}

\begin{abstract}
The decay of proton in the de Sitter environment is governed by the temperature $T=H/\pi$, where $H$ is the Hubble parameter. This temperature is twice larger than the Gibbons-Hawking temperature $T_{\rm GH}=H/2\pi$. This demonstrates the physical difference of two processes. The temperature $T=H/\pi$ determines the proton decay rate in the local process which takes well inside the cosmological horizon. While the Gibbons-Hawking  temperature can be related to the processes which involve the Hawking photons or other particles radiated from the cosmological horizon. The same temperature $T=H/\pi$ determines the radiation of electron-positron pairs by positron or by other object in the de Sitter environment. Creation of matter and its thermalization in the de Sitter heat bath leads to the energy exchange between matter and quantum vacuum and finally to the decay of the de Sitter state towards the Minkowski vacuum. The lesson for the Unruh effect is also discussed. One may expect the similar separation of effects governed by two different temperatures: the effects related to the Rindler horizon with the Unruh temperature $T_{\rm U}=a/2\pi$ and the local radiation with $T=a/\pi=2T_{\rm U}$.
\end{abstract}

\maketitle

\tableofcontents

\section{Introduction}

Hawking effect in black holes\cite{Hawking1974} and in the de Sitter spacetime\cite{GH1977} and also the Unruh effect (or Fulling-Davies-Unruh effect \cite{Fulling1973,Davis1975,Unruh1976}) are the cornerstone phenomena of quantum filed theory in curved spacetimes and in non-inertial reference frames.

The Unruh effect has been applied to the problem of the decay of the accelerated proton, $p^+ \rightarrow n + e^+ +\nu$, which is a kind of the inverse $\beta$-decay, see e.g. Refs.\cite{Matsas2001,Suzuki2002,Ahluwalia2016,Matsas2017,Blasone2020}. 
Decay of the non-inertial proton was discussed in the earlier papers, see Refs.\cite{Ginzburg1965,Ginzburg1965b} and references therein. Here we consider the similar proton decay in a different configuration -- in the de Sitter environment. 
The de Sitter spacetime has high symmetry due to constant in space and time scalar Riemann curvature.
This allows us to study different processes in the de Sitter state using its local thermodynamics.\cite{Volovik2024} 
 The de Sitter thermodynamics is characterized by the local temperature $T=H/\pi$, which is twice the Gibbons-Hawking temperature, $T=2T_{\rm GH}$. However, the holographic bulk-surface correspondence does work: the entropy integrated over the Hubble volume coincides with the Gibbons-Hawking entropy $A/4G$, where $A$ is the area of the cosmological horizon. 

Here we show that the same temperature $T=H/\pi$ governs the process of the proton decay in the de Sitter environment, and also the other processes generated by positron, such as electron-positron pair creation.
It fits into the overall scheme of the particle decay in the de Sitter environment. \cite{Bros2008,Volovik2009,Bros2010,Vanzo2011,Jatkar2012,Maldacena2015} 
The processes  electron-positron pair creation by proton (or by any other object) in the de Sitter environment demonstrates the instability of the de Sitter state and its power-law decay to the Minkowski vacuum.

The experience with proton in the de Sitter universe may provide some lessons for accelerated protons in Minkowski space. 

 \section{Decay of particles in de Sitter spacetime}
\label{DecayComposite}

 \subsection{Ionization of hydrogen atom}
\label{HydrogeneAtom}

In general, to consider the processes in the de Sitter environment it is convenient to use the metric in the Painlev\'e-Gullstrand (PG) form:\cite{Painleve,Gullstrand}
\begin{equation}
ds^2= - dt^2 +   (d{\bf r} - {\bf v}({\bf r})dt)^2\,.
\label{PG1}
\end{equation}
Here ${\bf v}({\bf r})$ is the shift velocity, which in the de Sitter state is ${\bf v}({\bf r})=H{\bf r}$, and the metric is:
\begin{equation}
ds^2= - dt^2 +   (dr - Hr dt)^2+r^2 d\Omega^2\,.
\label{PG}
\end{equation}
 
The PG metric is stationary, i.e. does not depend on time, and it does not have the unphysical singularity at the cosmological horizon. That is why it is appropriate for consideration of the local thermodynamics both inside and outside the horizon. This also allows us to discuss the processes related to the cosmological horizon at $r=r_h=1/H$. 

However, such processes as ionization of an atom take place well inside the cosmological horizon, at $r<r_h$, where the conventional static metric can be used:
\begin{equation}
ds^2= - (1-H^2r^2)dt^2 +   \frac{dr^2}{1-H^2r^2}+r^2 d\Omega^2\,.
\label{dSmetricStatic}
\end{equation}

The ionization of the hydrogen atom, $H\rightarrow p^+ + e^-$, is the typical example of splitting of the composite object in the de Sitter environment. The rate of this process of ionization can be calculated using the semiclassical tunneling picture.\cite{Volovik2024,Maxfield2022} The calculation of the imaginary part of the action gives the following activation exponent:
\begin{equation}
w \propto \exp(-2\,{\bf Im}\,S) = \exp\left(-\frac{\pi \epsilon}{H}\right)\,.
\label{activation}
\end{equation}
Here $\epsilon \gg H$ is ionization potential, and the semiclassical action is $S=\int dr \,p_r(r)$. The trajectory $p_r(r)$ of the radiated electron for $\epsilon\ll m_e$ is obtained from the classical non-relativistic equation 
\begin{equation}
\frac{{\bf p}^2}{2m_e}  - \frac{1}{2}m_eH^2r^2 = -\epsilon\,,
\label{GravPotential}
\end{equation}
which gives
\begin{eqnarray}
p_r(r)= \sqrt{m_e^2H^2r^2 -2m_e \epsilon}\,.
\label{ElectronTrajectory2}
\end{eqnarray}
The momentum $p(r)$ is imaginary at 
\begin{eqnarray}
r<r_0=\frac{1}{H}\sqrt{\frac{2\epsilon}{m_e}} \ll r_H=\frac{1}{H}\,.
\label{ImaginaryTrajectory}
\end{eqnarray}
Then one has 
\begin{eqnarray}
2\,{\bf Im}\,S= 2m_eH\int_0^{r_0}dr\sqrt{r_0^2 - r^2}=\frac{\pi\epsilon}{H}\,,
\label{ImaginaryTrajectory2}
\end{eqnarray}
 which gives Eq.(\ref{activation}).

In Eq.(\ref{activation}) we ignore the pre-exponential factor. In this approximation the decay rate coincides with that, which comes from the activation process in the Minkowski state in the presence of the heat bath with temperature $T=H/\pi$. This demonstrates that matter perceives the de Sitter environment as the heat bath with temperature $T=H/\pi$. The local temperature of the de Sitter heat bath governs also the other processes of  the splitting of stable objects, including the decay of proton in de Sitter environment. 
It is important that due to the de Sitter symmetry, the local temperature is the same for all points in the de Sitter space, i.e. for all observers, that are static in the de Sitter environment. The coordinate transformation in the PG metric (\ref{PG1}) which connects two static observers:
\begin{eqnarray}
{\bf r}\rightarrow {\bf r} + {\bf a}\,e^{Ht}\,\,,\,\, t\rightarrow t\,.
\label{TramsformationsObservers}
\end{eqnarray}
The metric (\ref{PG1}) is invariant under this shift of the position of the observer. Each observer considers the hydrogen atom, which is at rest in his/her reference frame.

Here we do not distinguish between composite and elementary (fundamental) particles (the so-called nuclear democracy\cite{Chew1964,Chew1966}). Both are considered as objects, which are stable in Minkowski vacuum and are unstable in de Sitter. In particular, we consider proton as a stable particle ignoring the possible proton decay to positron and neutral pion, which violates the conservation of baryon and lepton numbers.

\subsection{General case}
\label{General}

Keeping in mind the inverse $\beta$-decay, $p^+ \rightarrow n + e^+ +\nu$, we consider the general case, when the initial particle with mass $M_{\rm in}$ decays into $N$ particles with masses $m_a$ such that the total mass $M_{\rm fin}$ of the radiated particles is larger than $M_{\rm in}$:
\begin{equation}
M_{\rm in} \rightarrow M_{\rm fin}=\sum_{a=1}^N m_a >M_{\rm in} \,.
\label{Masses}
\end{equation}
As in the case of ionization of the hydrogen atom, such process is not allowed in the Minkowski vacuum, but is possible in the de Sitter environment or under acceleration. 

It is convenient to represent the initial particle as the collection of the $N$ virtual (off-shell) particles with energies $E_a<m_a$:
\begin{equation}
M_{\rm in} =\sum_a E_a \,.
\label{Masses}
\end{equation}
Since $E_a<m_a$, the radiation of each component is classically forbidden, but in the de Sitter environment it becomes possible due to the process of quantum tunneling (see also Ref.\cite{Vanzo2011}).
Then the process of decay of the initial particle into $N$ particles can be considered as $N$ simultaneous (coherent) processes, in which each virtual (off-shell) particle with energy $E_a$ transforms in the de Sitter environment to the on-shell particle with mass $m_a$.

The virtual particle becomes the real particle at some distance $r_0$ from the position $r=0$ of the initial  object. At this distance the gravitational potential of the de Sitter state compensates the difference between the energy $E_a$ of the virtual particle and the mass $m_a$ of the radiated particle, in the same way as this takes place in the process of ionization of the hydrogen atom.\cite{Volovik2024,Maxfield2022} 

Using the static de Sitter metric in Eq.(\ref{dSmetricStatic}) one obtains the trajectory of each particle moving in the radial direction from the origin at $r=0$ to the distance $r_0$: 
\begin{equation}
p_a(r) = \frac{1}{1-H^2r^2} \sqrt{ E_a^2 -m_a^2+ m_a^2H^2r^2}\,.
\label{ParticleTraj2}
\end{equation}
The momentum on this trajectory is imaginary in the classically forbidden region where the $a$-th particle is virtual, i.e. at $r<r_0$, where
\begin{equation}
r_0=\frac{\sqrt{1-\frac{E_a^2}{m_a^2}}}{H}\label{r0}  < r_H=\frac{1}{H}\,.
\end{equation}
In case of ionization of atom, one has $E_e=m_e -\epsilon$ for electron and $E_p=m_p$ for proton. 
For $\epsilon \ll m_e$ this gives $r_0=\sqrt{2\epsilon/m_e}/H$ in Eq.(\ref{ImaginaryTrajectory}).

 The imaginary momentum of the $a$-th particle gives the imaginary contribution to its action and thus the tunneling exponent:
\begin{equation}
{\bf Im}\,S_a={\bf Im}\int dr ~p_a(r)=\frac{m_a}{H}\int_0^{r_0}dr \frac{\sqrt{r_0^2-r^2}}{r_H^2-r^2} = \frac{\pi}{2} (m_a-E_a)\,.
\label{DecayExponent}
\end{equation}
Note that since $r_0<r_H$, the tunneling process takes place well inside the cosmological horizon and is not related to the process of Hawking radiation from the cosmological horizon. 

Now we must take into account that in the quantum process of splitting of the initial object, all $N$ virtual components tunnel simultaneously.  Such coherent process of simultaneous tunneling of several component (the co-tunneling) gives the following decay rate of the initial particle into $N$ particles (here we assume that $m_a\gg H$):
\begin{equation}
\Gamma(M_{\rm in} \rightarrow M_{\rm fin}) \propto \exp(-2\sum_a\,{\bf Im}\,S_a) =
 \exp\left(-\frac{\pi}{H}\sum_a(m_a-E_a)\right)= 
\exp\left(-\frac{\pi (M_{\rm fin} -M_{\rm in})}{H}\right)~.
\label{DecayRate2}
\end{equation}

This demonstrates that rate of the particle decay in the de Sitter environment in the limit $m_a\gg H$ is the same as in the heat bath with temperature $T=H/\pi$ in the Minkowski spacetime:
\begin{equation}
\Gamma(M_{\rm in} \rightarrow M_{\rm fin})  \sim  \exp{\left(-\frac{\Delta M}{T} \right)} \,\,,\,\, T=\frac{H}{\pi}\,.
\label{CompositeRate}
\end{equation}
Note that in the limit of small $H$ and ignoring the pre-exponential factor, which depends on the coupling constants of interactions responsible of decay, this result is universal. It does not depend on details of the process. In particular it does not depend on the distribution of energies $E_a$ between the virtual particles in Eq.(\ref{Masses}) and also on the number $N$ of the virtual states (a kind of nuclear democracy\cite{Chew1964,Chew1966}).

 \section{Proton in de Sitter environment}
\label{Proton}

 \subsection{Decay of proton in de Sitter environment}
\label{DecayProton}

Eq.(\ref{CompositeRate}) can be applied to the decay of proton in the inverse $\beta$-decay in the de Sitter environment:
\begin{equation}
p^+ \rightarrow n + e^+ +\nu\,.
\label{ProtonDecay}
\end{equation}
Eq.(\ref{CompositeRate}) for splitting into $N=3$ components gives the proton decay rate:
\begin{eqnarray}
\Gamma(p^+ \rightarrow n + e^+ +\nu) \sim  \exp{\left(-\frac{\pi\Delta M}{H} \right)}\,,
\label{ProtonDecayRate}
\\
\Delta M = M_{\rm fin} -M_{\rm in}= m_n +m_e + m_\nu -m_p\,.
\label{MassDifference}
\end{eqnarray}
Here for $m_\nu$ we can use the mass eigenstates for neutrino, where there is no distinction
between inertial and gravitational masses.\cite{Blasone2020b} 

In a similar way one can consider the following channel of decay of the hydrogen atom:
\begin{eqnarray}
\Gamma(H \rightarrow n + \nu) \sim  \exp{\left(-\frac{\pi\Delta M}{H} \right)}\,,
\label{AtomDecayRate}
\\
\Delta M = M_{\rm fin} -M_{\rm in}= m_n +m_\nu  -m_H\,.
\label{AtomMassDifference}
\end{eqnarray}

\subsection{Decay of proton in the Schwarzschild-de Sitter environment}
\label{SdSDecayProton}

We can use the conventional static metric for the Schwarzschild-de Sitter (SdS) black hole:
\begin{equation}
ds^2= - \left(1-\frac{2M}{r} -H^2r^2\right)dt^2 +   \frac{dr^2}{1-\frac{2M}{r} - H^2r^2}+r^2 d\Omega^2\,.
\label{dSmetricStatic}
\end{equation}
Let us place the proton at the point $r=r_0$, where
\begin{equation}
r_0^3=\frac{GM}{H^2}  \,.
\label{rest}
\end{equation}
The spherical surface at  $r=r_0$ is the zero-gravity surface, and thus 
we can find the proton decay rate, when proton is rest. Neat this point, at $|r-r_0|\ll r_0$, the metric is
\begin{equation}
ds^2= - \left(1-3H^2(r-r_0)^2\right)dt^2 +   \frac{dr^2}{1-3H^2(r-r_0)^2}+r^2 d\Omega^2\,.
\label{dSmetricPoint}
\end{equation}
The decay rate in such environment corresponds to the decay rate in the de Sitter state with the effective Hubble parameter
$\tilde H =\sqrt{3}H$. Then the temperature, which determines the decay of proton in the SdS environment, is (see also Ref. \cite{Volovik2023}):
\begin{equation}
T_{SdS}=\frac{\sqrt{3}}{\pi}H  \,.
\label{TSdS}
\end{equation}

It is interesting that this temperature does not depend on the mass of black hole and is fully determined by the Hubble parameter. 
The process of decay takes place in the region between the black hole horizon and the cosmological horizon, $r_-<r_0<r_+$, and 
thus it is the local process, which has no relation to the Hawking radiation from the horizons. However, in the limit when the two horizons merge we can compare this activation temperature with the Bousso-Hawking temperature.\cite{BoussoHawking1996} In this Nariai limit the Bousso-Hawking temperature $T_{\rm BH}$ is twice smaller than the temperature which determines the proton decay in the SdS spacetime, $T_{\rm BH}=\frac{\sqrt{3}}{2\pi}H=T_{SdS}/2$.

The quantum tunneling approach also determines the thermodynamics of black holes. The  quantum tunneling process of radiation of particles determines the Hawking temperature, while the quantum tunneling of macroscopic objects (the macroscopic quantum tunneling)  determines the non-extensive entropy of the black hole. \cite{Volovik2024TC}  

 \subsection{Proton as the source of electron-positron pairs}
\label{ProtonSource}

If the neutrino mass is neglected, the mass difference in proton decay in Eq.(\ref{MassDifference}) is:
\begin{equation}
\Delta M \approx m_e +(m_n -m_p)\approx 0.51 MeV + 1.29 MeV \approx 3.5 \, m_e > 2m_e\,.
\label{MassBalance}
\end{equation}
Since $\Delta M> 2m_e$, another process related to proton has the higher rate.  Since neutron is unstable, the proton in the de Sitter environment serves as the source of the electron-positron pair creation:
\begin{equation}
p^+ \rightarrow n + e^+ +\nu \rightarrow p^+ +e^++e^- + \nu + \bar\nu\,.
\label{PairCreation}
\end{equation}
The rate of the direct electron-positron pair creation process generated by proton (i.e. without the intermediate step of inverse $\beta$-decay) is:
\begin{equation}
\Gamma(p^+ \rightarrow p^+ +e^++e^-) \sim  \exp{\left(-\frac{2\pi m_e}{H} \right)} = \exp{\left(-\frac{2m_e}{T} \right)}\,.
\label{PairCreationRate}
\end{equation}
Here we consider the process in which neutrino and antineutrino are either annihilated or are not formed.
In the latter case the virtual energies in Eq.(\ref{Masses}) are $E_{e^+}=E_{e^-}=0$, while proton is on-shell, $E_{p^+}=m_p$.

Note that in the Gibbons-Hawking radiation from the cosmological horizon, with such rate only a single electron (or a single positron) can be created, $w\propto \exp(-m_e/T_{GH})$. For the coherent creation of pairs, the double Gibbons-Hawking temperature  $T=2T_{GH}$ is needed.  The reason for such difference is that in our case the process of pair creation takes place well inside the cosmological horizon. It is generated by the positron: its presence in the de Sitter vacuum violates the de Sitter symmetry and allows for the creation of pairs from the vacuum by providing the non-zero matrix elements. This demonstrates the physical difference between the local processes governed by temperature $T=H/\pi$ and the processes generated by the cosmological horizon with the Gibbons-Hawking temperature $T_{GH}=H/2\pi$, which is by the factor 2 smaller.

The analogous process of creation of the  neutrino-antineutrino pairs generated by proton in the de Sitter environment has the rate:
\begin{equation}
\Gamma(p^+ \rightarrow p^+ + \nu + \bar\nu) \sim  \exp{\left(-\frac{2\pi m_\nu}{H} \right)} \,.
\label{PairNeutrinoRate}
\end{equation}

 \section{Vacuum back reaction and de Sitter decay}
\label{BackReaction_sec}

The process of creating electron-positron and neutrino-antineutrino pairs by a positron in equations (\ref{PairCreationRate}) and (\ref{PairNeutrinoRate}) can be repeated many times.  The same positron inserted into the de Sitter vacuum may create more and more pairs. This demonstrates the instability of the de Sitter vacuum with respect to the creation of matter if a positron or other external object is present in the de Sitter Universe. This instability has no relation to the cosmological horizon. It is initiated by the external objects. These external objects serve as sources of particle radiation in the de Sitter medium, while the emitted particles in turn become sources of further radiation. 

This leads to the following consequences. First, since the processes of radiation are determined by the characteristic temperature $T=H/\pi$ of the de Sitter heat bath, one may expect that the temperature of the radiated matter will tend to approach the temperature of the heat bath. Second, the creation of matter with its energy density must be followed by the back reaction of the vacuum, which looses energy.

The decay of the quantum vacuum due to activation and then the thermalization of matter can be considered using the extension of the Starobinsky approach.\cite{Starobinsky1977,Starobinsky1982,Starobinsky1986,Starobinsky1994,Starobinsky1996,Starobinsky1997,Dobado1999,Starobinsky2023} 
Here we consider the simple phenomenological scenario\cite{Volovik2024} based on the energy exchange between the de Sitter heat bath and the thermal matter under the slow-roll condition, $|\dot H| \ll H^2$.  Without the back reaction, the Friedmann equations give the following decay of matter density due to the Hubble friction, 
\begin{eqnarray}
\partial_t \rho_M= - 3(1+w)H \rho_M\,.
\label{MatterConservation}
\end{eqnarray}
This equation is modified, when we take into account the heat exchange. We consider here the case when matter is represented by the  relativistic gas of massless particles, i.e. $P_M=w \rho_M$ with $w=1/3$.

Since the relativistic gas tends to approach  the temperature $T=H/\pi$ of the de Sitter environment, its energy density   $\rho_M$ tends to approach the value $\rho_M(H)=bH^4$, where  the dimensionless parameter $b$ depends on the number of massless relativistic fields. Then the natural modification of Eq.(\ref{MatterConservation}) is:
\begin{eqnarray}
\partial_t \rho_M= - 4H (\rho_M -\rho_M(H))\,\,,\,\,\rho_M(H)=bH^4 \,.
\label{MatterNonConservation}
\end{eqnarray}
The extra gain of the matter energy, $4H\rho_M(H)$, must be compensated by the corresponding loss of the vacuum energy:
\begin{eqnarray}
\partial_t \rho_{\rm vac}= - 4H\rho_M(H) \,.
\label{VacuumNonConservation}
\end{eqnarray}

Since the vacuum energy density is $\rho_{\rm vac} =3H^2/8\pi G$, one obtains from Eq.(\ref{VacuumNonConservation}) the following time dependence of the Hubble parameter and of the energy densities of vacuum and matter:
 \begin{eqnarray}
H(t) =b^{-1/3} M_{\rm Pl} \left( \frac{t_{\rm Pl}}{t+t_0}\right)^{1/3}  \,,
\label{DecayLawH}
\\
\rho_{\rm vac}(t) =6b^{-2/3} M_{\rm Pl}^4  \left( \frac{t_{\rm Pl}}{t+t_0}\right)^{2/3} \,.
\label{DecayLawV}
\end{eqnarray}
Here $M_{\rm Pl}^2=1/16\pi G$ and $t_{\rm Pl}=1/M_{\rm Pl}$, and we assume that $t_0 \gg t_{\rm Pl}$, and thus $|\dot H| \ll H^2$.

 The power law $1/3$ of decay of $H$ in Eq.(\ref{DecayLawH}) agrees with the results obtained using several different approaches. \cite{Padmanabhan2003,Padmanabhan2005,KlinkhamerVolovik2016,Markkanen2018,Markkanen2018a,Roman2020,Gong2021}
This demonstrates that the phenomenological scenario of thermalization of matter by the de Sitter heat bath in Eqs. (\ref{MatterNonConservation}) and (\ref{VacuumNonConservation}) is reasonable.  This scenario  does not require any microscopic or macroscopic model and describes the Starobinsky inflation in terms of the phenomenological parameters $t_0$. This parameter determines the initial value of the Hubble parameter at the beginning of inflation:
 $H(t=0) =b^{-1/3}  M_{\rm Pl} \left(\frac{t_{\rm Pl}}{t_0}\right)^{1/3} \ll M_{\rm Pl}$, which corresponds to the scaleron mass in Starobinsky inflation.

In this approach there is no cosmological constant problem. Even if the initial vacuum energy is on the order of the Planck scale, the vacuum looses its energy and finally decays towards the perfect vacuum -- the Minkowski vacuum. The energy of the stable Minkowski vacuum is always zero,\cite{KlinkhamerVolovik2008} as it follows from the thermodynamic Gibbs-Duhem relation, $\rho=-P=0$, which is applicable to the ground state of any macroscopic system (relativistic and non-relativistic) in the absence of external environment.

On the other hand, the pure de Sitter vacuum can be considered as the combination of dark energy with the equation of state  $w=-1$ and dark matter with the equation of state $w=+1$.\cite{Volovik2024ad}  The dark matter component of the de Sitter vacuum has the same equation of state as the Zel'dovich stiff matter.\cite{Zeldovich1962}

\section{Discussion}
\label{DiscussionSec}

We considered the decay of proton in the de Sitter environment. What can be the lessons for the decay of the accelerated proton considered in Refs.\cite{Matsas2001,Suzuki2002,Ahluwalia2016,Matsas2017,Blasone2020}? 

One lesson is that there are two different temperatures, which govern the decay of stable particles in the de Sitter environment: the local temperature $T=H/\pi$ and the twice smaller Gibbons-Hawking temperature $T_{\rm GH}=H/2\pi$ related to the cosmological horizon. The local temperature determines the proton decay rate in the local processes which occur well inside the cosmological horizon. On the other hand, the Gibbons-Hawking temperature can be related to the processes which involve the Hawking photons, electrons and other particles radiated from the cosmological horizon. 

The similar division may take place for the accelerated protons. There can be the local processes and the processes related to the Rindler horizon. For example, the doubling of the Unruh temperature takes place in the particular case of the combined Schwinger-Unruh process.\cite{Volovik2024a} This is similar to the doubling of the Gibbons-Hawking temperature.
This demonstrates that the FDU phenomenon is rather delicate giving rise to some controversies or requires some modifications.\cite{Ford2006,Blasone2021,Gregori2024} In particular, it depends on the way how the observer measures the temperature, and on his/her ability to get the full information about the process. 

Another lesson from de Sitter is that in the de Sitter environment the proton radiates electron-positron pairs. The rate of the emission of pairs is also determined by the local temperature $T=H/\pi$. This is distinct from the Hawking radiation of electrons and positrons caused by the cosmological horizon, which is determined by the  twice smaller Gibbons-Hawking temperature. One may expect the similar doubling of the Unruh temperature in the radiation of pairs by accelerated proton. If so this would be in favour of the separation of two effects with two different temperatures: the effect related to the Rindler horizon with Unruh temperature $T_{\rm U}=a/2\pi$ and the local radiation with $T=a/\pi=2T_{\rm U}$.

It would be interesting to consider, how these two sides of thermodynamics (the local thermodynamics and the thermodynamics related to the cosmological and Rindler horizons) are modified in the time-dependent environment. The particular case is when the de Sitter state is perturbed towards the Friedmann–Lemaitre–Robertson–Walker (FLRW) cosmology.\cite{Odintsov2024}
Such transformation from de Sitter to FLRW takes place due to processes of creation of electron-positron pairs initiated by proton or by any other object inserted into the de Sitter Universe.  These processes lead to the power-law decay of the de Sitter state towards the Minkowski vacuum due to the energy exchange between the created matter and the de Sitter heat bath.

{\bf Acknowledgements}. I thank Oleg Andreev for discussions.

\end{document}